\documentclass[preprint2]{aastex}
\usepackage{txfonts}
\usepackage{epsfig}
\usepackage{graphicx}
\usepackage{natbib}

\shorttitle{A seismic signature of a second dynamo?}
\shortauthors{Fletcher et al.}
\begin{document}

\title{A seismic signature of a second dynamo?}

\author{\textsc{Stephen T. Fletcher$^1$, Anne-Marie Broomhall$^2$,
David Salabert$^{3,}$\altaffilmark{6}, Sarbani Basu$^4$, William J.
Chaplin$^2$, Yvonne Elsworth$^2$, \large Rafael A. Garcia$^5$, and
Roger New$^1$}}\affil{\footnotesize $^1$Faculty of Arts, Computing,
Engineering and Sciences, Sheffield Hallam University, Sheffield S1
1WB, UK;
S.Fletcher@shu.ac.uk, R.New@shu.ac.uk\\
$^2$School of Physics and Astronomy, University of Birmingham,
Edgbaston, Birmingham B15 2TT, UK;\\ amb@bison.ph.bham.ac.uk,
wjc@bison.ph.bham.ac.uk, ype@bison.ph.bham.ac.uk\\
$^3$Instituto de Astrof{\'{i}}sica de Canarias, E-38200 La Laguna,
Tenerife, Spain; salabert@iac.es\\ $^4$Yale University, P.O. Box
208101, New Haven, CT 06520-8101, USA; sarbani.basu@yale.edu\\
$^5$Laboratoire AIM, CEA/DSM-CNRS-Universit{\'e} Paris Diderot,
IRFU/SAp, Centre de Saclay, 91191 Gif-sur-Yvette, France;
rafael.garcia@cea.fr}

\altaffiltext{6}{Departamento de Astrof{\'{i}}sica, Universidad de
La Laguna, E-38205 La Laguna, Tenerife, Spain}

\begin{abstract}The Sun is a variable star whose magnetic activity
varies most perceptibly on a timescale of approximately 11 years.
However, significant variation is also observed on much shorter
timescales. We observe a quasi-biennial (2 year) signal in the
natural oscillation frequencies of the Sun. The oscillation
frequencies are sensitive probes of the solar interior and so by
studying them we can gain information about conditions beneath the
solar surface. Our results point strongly to the 2 year signal being
distinct and separate from, but nevertheless susceptible to the
influence of, the main 11 year solar cycle.
\end{abstract}


\section{Introduction}\label{section[introduction]}
The Sun is a variable star, whose magnetic activity shows systematic
variations. The most conspicuous of these variations is the 11 year
solar cycle \citep{Hathaway2010}. Starting in 1755 March, the 11
year cycles have been labeled with a number and we are currently
moving into cycle 24 after just coming out of an unusually extended
minimum.

Over the past twenty years it has become apparent that significant
(quasi-periodic) variability is also seen on shorter timescales,
between 1 and 2 years \citep[e.g.][]{Benevolenskaya1995,Mursula2003,
Valdes-Galicia2008}. In this Letter we investigate the origins of
this so-called ``mid-term'' periodicity by looking beneath the solar
surface. We attempt to answer the following question: Is the
periodicity the result of modulation of the main solar dynamo
responsible for the 11 year cycle or is it caused by a separate
mechanism?

%%%%%%%%%%%%%%%%%%%%%%%%%%%%%%%%%%%%%%%%%%%%%%%%%%%%%%%%%%%%%%%%%%%%
\begin{figure*}
  \centering
  \includegraphics[width=0.7\textwidth, clip]{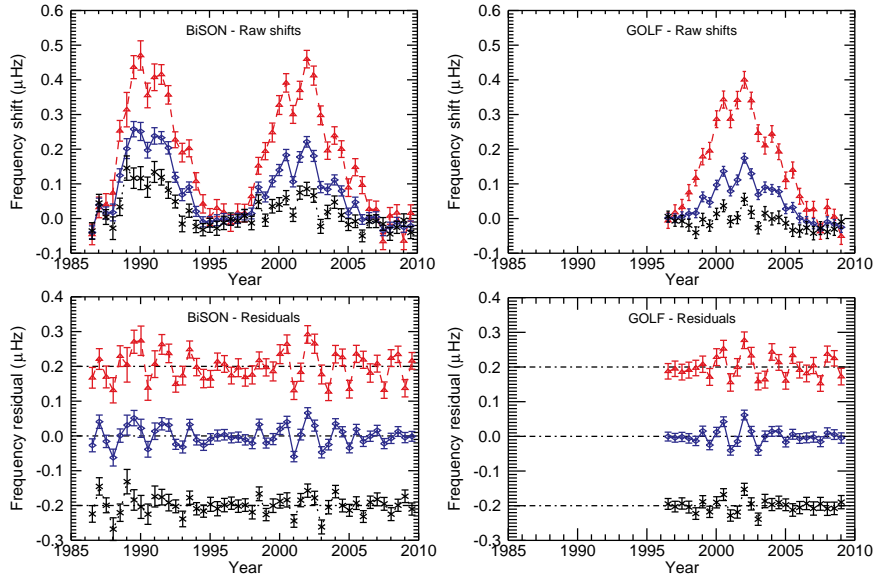}\\
  \caption{\small{Top: average frequency shifts of ``Sun-as-a-star'' modes with
  frequencies between 1.88 and $3.71\,\rm mHz$ (total-frequency band, blue solid line, and diamond symbols); 1.88 and $2.77\,\rm
  mHz$ (low-frequency band, black dotted line, and cross symbols); and 2.82 and $3.71\,\rm mHz$
  (high-frequency band, red dashed line, and triangle symbols). Bottom:
residuals left after dominant 11 year signal has been removed (black
dotted and red dashed curves are displaced by $-0.2$ and $+0.2$,
respectively, for clarity).}}\label{figure[results]}
\end{figure*}
%%%%%%%%%%%%%%%%%%%%%%%%%%%%%%%%%%%%%%%%%%%%%%%%%%%%%%%%%%%%%%%%%%%%

To answer this question we have used helioseismology to investigate
the solar-cycle-related changes of the Sun's interior. We study the
variation with the solar cycle of the frequencies of the Sun's
natural modes of oscillation, which are known as p modes (because
internal gradients of pressure provide the restoring force for the
oscillations). Solar p modes are trapped in cavities below the
surface of the Sun; and their frequencies are sensitive to
properties, such as temperature and mean molecular weight, of the
solar material in the cavities \citep[e.g.][and references
therein]{JCD2002}. The frequencies of p modes vary throughout the
solar cycle with the frequencies being at their largest when the
solar activity is at its maximum \citep[e.g.][]{Woodard1985,
Palle1989, Elsworth1990, Jimenez2003, Chaplin2007, Jimenez2007}. By
examining the changes in the observed p-mode frequencies throughout
the solar cycle, we can learn about solar-cycle-related processes
that occur beneath the Sun's surface.

We use oscillations data collected by making unresolved
(Sun-as-a-star) Doppler velocity observations, which are sensitive
to the p modes with the largest horizontal scales (or the lowest
angular degrees, $\ell$). Consequently, the observed frequencies are
of the truly global modes of the Sun \citep[e.g.][and references
therein]{JCD2002}. These modes travel to the Sun's core but, because
the sound speed inside the Sun increases with depth, their dwell
time at the surface is longer than at the solar core. Consequently,
p modes are most sensitive to variations in regions of the interior
that are close to the surface and so are able to give a global
picture of the influence of near-surface activity. The data were
collected by the Birmingham Solar-Oscillations Network
\citep[BiSON;][]{Elsworth1995, Chaplin1996} and the Global
Oscillations at Low Frequencies \citep[GOLF;][]{Gabriel1995,
Jimenez2003, Garcia2005} instrument on board the ESA/NASA
\emph{Solar and Heliospheric Observatory} (\emph{SOHO}) spacecraft.

%%%%%%%%%%%%%%%%%%%%%%%%%%%%%%%%%%%%%%%%%%%%%%%%%%%%%%%%%%%%%%%%%%%%
\begin{figure}
  \centering
  \includegraphics[width=0.9\columnwidth, clip]{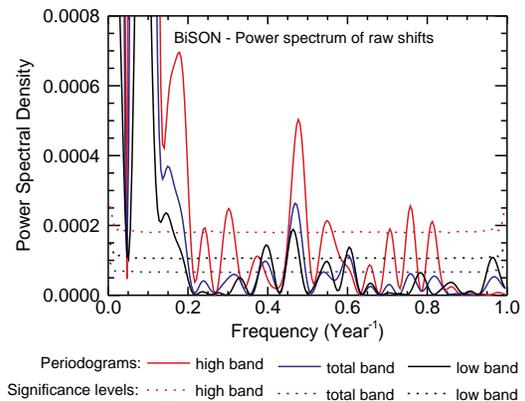}\\
  \caption{\small{Periodograms of the BiSON frequency shifts. The solid lines
  represent the periodograms of the different frequency bands (see legend). The dotted lines represent the
  1\%
  ``false alarm'' significance levels
  for the respective frequency ranges.}}\label{figure[periodogram]}
\end{figure}
%%%%%%%%%%%%%%%%%%%%%%%%%%%%%%%%%%%%%%%%%%%%%%%%%%%%%%%%%%%%%%%%%%%%

\section{Uncovering the mid-term
periodicity}\label{section[results]}

The observations made by BiSON and GOLF were divided into
182.5-day-long independent subsets. BiSON has now been collecting
data for over 30 years. The quality of the early data, however, is
poor compared to more recent data because of poor time coverage.
Here, we have analyzed the mode frequencies observed by BiSON during
the last two solar cycles in their entirety i.e. from 1986 April 14
to 2009 October 7. GOLF has been collecting data since 1996 and so
we have been able to analyze data covering almost the entirety of
solar cycle 23, i.e., from 1996 April 11 to 2009 April 7. After 1996
April 11, when both GOLF and BiSON data were available, we ensured
that the start times of the BiSON and GOLF subsets were the same.
Note that at the time of writing we have more recent calibrated data
available from BiSON than from GOLF.

Estimates of the mode frequencies were extracted from each subset by
fitting a modified Lorentzian model to the data using a standard
likelihood maximization method. Two different fitting codes have
been used to extract the mode frequencies, both giving the same
results. For clarity, we only show the results of one method, which
was applied in the manner described in \citet{Fletcher2009}. A
reference frequency set was determined by averaging the frequencies
in subsets covering the minimum activity epoch at the boundary
between cycle 22 and cycle 23. It should be noted that the main
results of this Letter are insensitive to the exact choice of
subsets used to make the reference frequency set. Frequency shifts
were then defined as the differences between frequencies given in
the reference set and the frequencies of the corresponding modes
observed at different epochs \citep{Broomhall2009}.

For each subset in time, three weighted-average frequency shifts
were generated, where the weights were determined by the formal
errors on the fitted frequencies: first, a ``total'' average shift
was determined by averaging the individual shifts of the $\ell=0$,
1, and 2 modes over fourteen overtones (covering a frequency range
of $1.88-3.71\,\rm mHz$); second, a ``low-frequency'' average shift
was computed by averaging over seven overtones whose frequencies
ranged from 1.88 to $2.77\,\rm mHz$; and third, a ``high-frequency''
average shift was calculated using seven overtones whose frequencies
ranged from 2.82 to $3.71\,\rm mHz$. The lower limit of this
frequency range (i.e., $1.88\,\rm mHz$) was determined by how low in
frequency it was possible to accurately fit the data before the
modes were no longer prominent above the background noise. The upper
limit on the frequency range (i.e., $3.71\,\rm mHz$) was determined
by how high in frequency the data could be fitted before errors on
the obtained frequencies became too large due to increasing line
widths causing modes to overlap in frequency.

The top panels of Figure \ref{figure[results]} show mean frequency
shifts of the p modes observed by BiSON and GOLF, respectively
\citep[also see][]{Broomhall2009, Salabert2009}. The 11 year cycle
is seen clearly and its signature is most prominent in the
higher-frequency modes. This is a telltale indicator that the
observed 11 year signal must be the result of changes in acoustic
properties in the few hundred kilometers just beneath the visible
surface of the Sun, a region that the higher-frequency modes are
much more sensitive to than their lower-frequency counterparts
because of differences in the upper boundaries of the cavities in
which the modes are trapped \citep{Libbrecht1990, JCD1991}. Despite
the low- and high-frequency bands showing different sensitivities to
the 11 year cycle there is a significant correlation between the
observed frequency-shifts. The correlations between the low- and
high-frequency band shifts are 0.82 for the BiSON data and 0.67 for
the GOLF data. The errors on the correlations indicate that there is
less than a 0.05\% chance that each of these correlations would
occur by chance.

In order to extract mid-term periodicities, we subtracted a smooth
trend from the average total shifts by applying a boxcar filter of
width 2.5 years. This removed the dominant 11-year signal of the
solar cycle. Note that, although the width of this boxcar is only
slightly larger than the periodicity we are examining here, wider
filters produce similar results. The resulting residuals, which can
be seen in the bottom panels of Figure \ref{figure[results]}, show a
periodicity on a timescale of about 2 years.

The signal is reassuringly similar in the BiSON and GOLF data sets.
The correlation between the two sets of frequency shifts was found
to be highly significant in all three frequency bands: 0.71 for the
low-frequency band, 0.99 for the high-frequency band, and 0.96 for
the total-frequency band. There is less than a 0.05\% chance that
these correlations would occur randomly. There is also a significant
correlation between the low- and high-frequency band residuals for
both the BiSON (0.46) and GOLF (0.55) data and there is less than a
1\% probability of these correlations occurring by chance.

Periodograms of the raw frequency shifts, as shown in the upper
panels of Figure \ref{figure[results]}, were computed to assess the
significance of the 2 year signal. Figure \ref{figure[periodogram]}
shows the periodograms obtained from the BiSON data, oversampled by
a factor of 10. The large peak at $0.09\,\rm yr^{-1}$ is the signal
from the 11-year cycle. There are also large peaks at approximately
$0.5\,\rm yr^{-1}$. Statistical analysis of the BiSON periodograms
established that the apparent 2-year periodicity was indeed
significant, in both the low- and high-frequency bands, with a false
alarm probability of 1\% \citep{Chaplin2002}. Excess power is also
present in the GOLF periodograms around $0.5\,\rm yr^{-1}$. However,
none of the GOLF peaks are as prominent as the equivalent peaks
observed in the BiSON data because fewer GOLF data are available,
particularly during periods of high activity when the 2 year signal
is most prominent.

\section{Discussion}\label{section[discussion]} The 2-year signal is
most prominent during periods that coincide with maxima of the 11
year cycle (although the high-frequency band continues to show
changes through the most recent solar minimum). But what is most
remarkable is that the 2 year signal has a similar amplitude in
modes of all frequencies. This is in stark contrast to the 11 year
signal, which is about five times stronger in the higher-frequency
modes (see Figure \ref{figure[results]}).

The acoustic 2 year signal we see in the modes is, therefore,
predominantly an \emph{additive} contribution to the acoustic 11
year signal; nevertheless, its amplitude envelope appears to be
modulated by the 11 year cycle. The 2 year signal must have its
origin in significantly deeper layers than the 11 year signal. Since
the 2 year signal shows far less dependence on mode frequency, the
origin of the signal must be positioned below the upper turning
point of the lowest frequency modes examined (as the depth of a
mode's upper turning point increases with decreasing frequency). The
upper turning point of modes with frequencies of 1.88\,mHz occurs at
a depth of approximately 1000\,km, whereas the influence of the 11
year cycle is concentrated in the upper few 100\,km of the solar
interior. Put together, this all points to a phenomenon that is
separate from, but nevertheless susceptible to, the influence of the
11 year cycle.

One possibility is a dynamo action seated near the bottom of the
layer extending 5\% below the solar surface. This region shows
strong rotational shear, like the shear observed across the
deeper-seated tachocline where the omega effect of the main dynamo
is believed to operate \citep{Corbard2002, Antia2008}. The presence
of two different types of dynamo operating at different depths has
already been proposed to explain the quasi-biennial behavior
observed in other proxies of solar activity
\citep{Benevolenskaya1998a, Benevolenskaya1998b}. When the 11 year
cycle is in a strong phase, buoyant magnetic flux sent upward from
the base of the envelope by the main dynamo could help to nudge flux
processed by this second dynamo into layers that are shallow enough
to imprint a detectible acoustic signature on the modes. The 2 year
signal would then be visible. When the main cycle is in a weak
phase, the flux from the second dynamo would not receive an extra
nudge, and would not be buoyant enough to be detected in the modes,
or in other proxies (where the 2 year signals are also only
detectible during phases of high activity) \citep[see, e.g.][and
references therein]{Vecchio2008, Hathaway2010}. That the signal was
also visible in the high-frequency modes during the recent extended
minimum may also, therefore, point to behavioral changes in the main
dynamo and its influence on the 2-year signal.

\acknowledgements This Letter utilizes data collected by the
Birmingham Solar-Oscillations Network (BiSON), which is funded by
the UK Science Technology and Facilities Council (STFC). We thank
the members of the BiSON team, colleagues at our host institutes,
and all others, past and present, who have been associated with
BiSON. The GOLF instrument on board \emph{SOHO} is a cooperative
effort of many individuals, to whom we are indebted. \emph{SOHO} is
a project of international collaboration between ESA and NASA.
S.T.F., A.M.B., W.J.C., Y.E., and R.N. acknowledge the financial
support of STFC. S.B. acknowledges NSF grants ATM-0348837 and
ATM-0737770. D.S. acknowledges the support of the grant
PNAyA2007-62650 from the Spanish National Research Plan. R.A.G.
thanks the support of the CNES/GOLF grant at the CEA/Saclay.

\end{document}